\documentclass[journal]{IEEEtran}

\usepackage{graphicx}
\usepackage{multirow}
\usepackage{amssymb}
\usepackage{amsmath}
\usepackage{mathrsfs}
\usepackage{booktabs}
\usepackage{fancyhdr}
\usepackage{lipsum}
\usepackage[T1]{fontenc}
\usepackage{mathtools}
\usepackage[final]{microtype}
\usepackage[space]{cite}
\usepackage{color}

\ifCLASSINFOpdf
\else
\fi 

\title{Weakly Labelled AudioSet Tagging\\with Attention Neural Networks}

\author{Qiuqiang Kong,~\IEEEmembership{Student Member,~IEEE}, Changsong Yu, Yong Xu,~\IEEEmembership{Member,~IEEE}, Turab Iqbal, \\ Wenwu Wang,~\IEEEmembership{Senior Member,~IEEE} and Mark D. Plumbley,~\IEEEmembership{Fellow,~IEEE}\thanks{Manuscript received March 13, 2019; revised June 18, 2019; accepted July 11, 2019. Date of publication July 26, 2019; date of current version August 21, 2019. This work was supported in part by the EPSRC Grant EP/N014111/1 ``Making Sense of Sounds'', in part by the Research Scholarship from the China Scholarship Council 201406150082, and in part by a studentship (Reference: 1976218) from the EPSRC Doctoral Training Partnership under Grant EP/N509772/1. The associate editor coordinating the review of this manuscript and approving it for publication was Dr. Alexey Ozerov. (\textit{Qiuqiang Kong is first author.) (Corresponding author: Yong Xu.)}}
\thanks{Q. Kong, T. Iqbal, and M. D. Plumbley are with the Centre for Vision, Speech and Signal Processing, University of Surrey, Guildford GU2 7XH, U.K. (e-mail: q.kong@surrey.ac.uk; t.iqbal@surrey.ac.uk; m.plumbley@surrey.ac.uk).}
\thanks{Y. Xu is with the Tencent AI Lab, Bellevue, WA 98004 USA (e-mail:
lucayongxu@tencent.com).}
\thanks{W. Wang is with the Centre for Vision, Speech and Signal Processing,
University of Surrey, Guildford GU2 7XH, U.K., and also with Qingdao
University of Science and Technology, Qingdao 266071, China (e-mail:
w.wang@surrey.ac.uk).}
\thanks{Digital Object Identifier 10.1109/TASLP.2019.2930913}
}

\begin{document}

\maketitle
\thispagestyle{fancy}

\begin{abstract}
Audio tagging is the task of predicting the presence or absence of sound classes within an audio clip. Previous work in audio tagging focused on relatively small datasets limited to recognising a small number of sound classes. We investigate audio tagging on AudioSet, which is a dataset consisting of over 2 million audio clips and 527 classes. AudioSet is weakly labelled, in that only the presence or absence of sound classes is known for each clip, while the onset and offset times are unknown. To address the weakly-labelled audio tagging problem, we propose attention neural networks as a way to attend the most salient parts of an audio clip. We bridge the connection between attention neural networks and multiple instance learning (MIL) methods, and propose decision-level and feature-level attention neural networks for audio tagging. We investigate attention neural networks modelled by different functions, depths and widths. Experiments on AudioSet show that the feature-level attention neural network achieves a state-of-the-art mean average precision (mAP) of 0.369, outperforming the best multiple instance learning (MIL) method of 0.317 and Google's deep neural network baseline of 0.314. In addition, we discover that the audio tagging performance on AudioSet embedding features has a weak correlation with the number of training samples and the quality of labels of each sound class. 

\end{abstract}

\begin{IEEEkeywords}
Audio tagging, AudioSet, attention neural network, weakly labelled data, multiple instance learning. 
\end{IEEEkeywords}

\IEEEpeerreviewmaketitle

\section{Introduction}
Audio tagging is the task of predicting the tags of an audio clip. Audio tagging is a multi-class tagging problem to predict zero, one or multiple tags for an audio clip. As a specific task of audio tagging, audio scene classification often involves the prediction of only one label in an audio clip, i.e. the type of environment in which the sound is present. In this paper, we focus on audio tagging. Audio tagging has many applications such as music tagging \cite{fu2011survey} and information retrieval \cite{typke2005survey}. An example of audio tagging that has attracted significant attention in recent years is the classification of environmental sounds, that is, predicting the scenes where they are recorded. For instance, the Detection and Classification of Acoustic Scenes and Events (DCASE) challenges \cite{dcase2013, dcase2015, dcase2016, mesaros2017dcase} consist of tasks from a variety of domains, such as DCASE 2018 Task 1 classification of outdoor sounds, DCASE 2017 Task4 tagging of street sounds and DCASE 2016 Task4 tagging of domestic sounds. These challenges provide labelled datasets, so it is possible to use supervised learning algorithms for audio tagging. However, many audio tagging datasets are relatively small \cite{dcase2013, dcase2015, dcase2016, mesaros2017dcase}, ranging from hundreds to thousands of training samples, while modern machine learning methods such as deep learning \cite{deep-learning-lecun, deep-learning-schmidhuber} often benefit greatly from larger dataset for training. 

In 2017, a large-scale dataset called \textit{AudioSet} \cite{audioset} was released by Google. AudioSet consists of audio clips extracted from YouTube videos, and is the first dataset that achieves a similar scale to the well-known ImageNet \cite{deng2009imagenet} dataset in computer vision. The current version (v1) of AudioSet consists of 2,084,320 audio clips organised into a hierarchical ontology with 527 predefined sound classes in total. Each audio clip in AudioSet is approximately 10 seconds in length, leading to 5800 hours of audio in total. AudioSet provides an opportunity for researchers to investigate a large and broad variety of sounds instead of being limited to small datasets with limited sound classes. 

\begin{figure}[t]
  \centering
  \centerline{\includegraphics[width=\columnwidth]{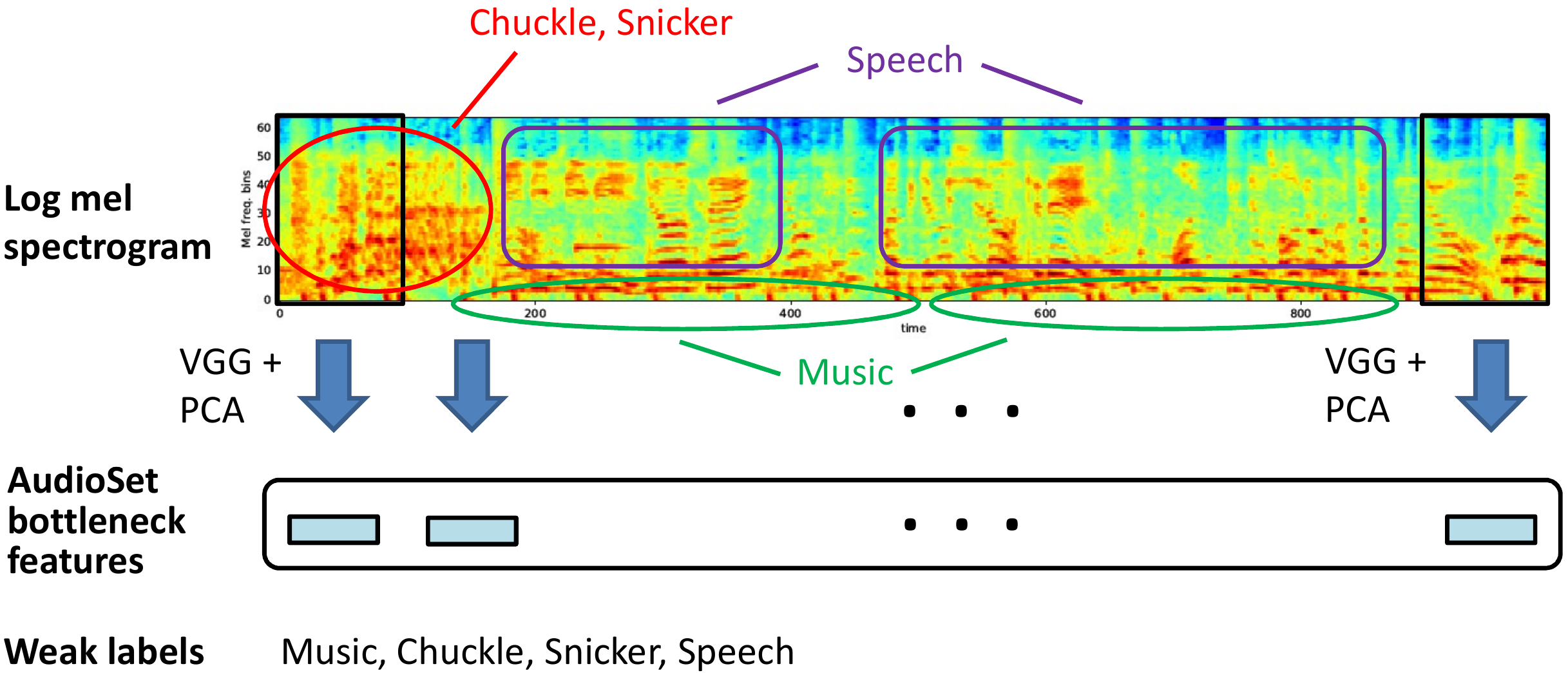}}
  \caption{From top to bottom: Log mel spectrogram of a 10-second audio clip; AudioSet bottleneck features extracted by a pre-trained VGGish convolutional neural network followed by a principle component analysis (PCA) \cite{hershey2017cnn}; Weak labels of the audio clip. There are no onset and offset times of the sound classes. }
  \label{fig:wav}
\end{figure}

One challenge of AudioSet tagging is that AudioSet is a weakly-labelled dataset (WLD) \cite{kumar2016audio, kong2017joint}. That is, for each audio clip in the dataset, only the presence or the absence of sound classes is indicated, while the onset and offset times are unknown. In previous work in audio tagging, an audio clip is usually split into segments and each segment is assigned with the label of the audio clip \cite{kong2016deep}. However, as the onset and offset of sound events are unknown so such label assignment can be incorrect. For example, a transient sound event may only appear a short time in a long audio recording. The duration of sound events can be very different and there is no prior knowledge of their duration. Different from ImageNet \cite{deng2009imagenet} for image classification where objects are usually centered and have similar scale, in AudioSet the duration of sound events may vary a lot. To illustrate, Fig. \ref{fig:wav} from top to bottom shows: the log mel spectrogram of a 10-second audio clip\footnote{https://www.youtube.com/embed/Wxa36SSZx8o?start=70\&end=80}; AudioSet bottleneck features \cite{audioset} extracted by a pre-trained VGGish convolutional network followed by a principal component analysis (PCA); weak labels of the audio including ``music'', ``chuckle'', ``snicker'' and ``speech'. In contrast to WLD, strongly labelled data (SLD) refers to the data labelled with both the presence of sound classes as well as their onset and offset times. For example, the sound event detection tasks in DCASE challenge 2013, 2016, 2017 \cite{dcase2013, dcase2016, mesaros2017dcase} provide SLD. However, labelling onset and offset times of sound events is time-consuming, so these strongly labelled datasets are usually limited to a relatively small size \cite{dcase2013, dcase2016, mesaros2017dcase}, which may limit the performance of deep neural networks that require large data to train a good model. 

In this paper, we train an audio tagging system on the large-scale weakly labelled AudioSet. We bridge our previously proposed attention neural networks \cite{kong2017audio, yu2018multi} with multiple instance learning (MIL) \cite{maron1998framework} and propose decision-level and feature-level attention neural networks for audio tagging. The contributions of this paper include the following:
\begin{itemize}
  \item Decision-level and feature-level attention neural networks are proposed for audio tagging; 
  \item Attention neural networks modelled by different functions, widths and depth are investigated; 
  \item The impact of the number of training samples per class on the audio tagging performance is studied; 
  \item The impact of the quality of labels on the audio tagging performance is studied. 
\end{itemize}

This paper is organised as follows. Section II introduces audio tagging with weakly labelled data. Section III introduces our previously proposed attention neural networks \cite{kong2017audio, yu2018multi}. Section IV introduces multiple instance learning. Section V reviews attention neural networks under the MIL framework and proposes decision-level and feature-level attention models. Section VI shows the experimental results. Section VII concludes and forecasts future work. 

\section{Audio Tagging with weakly labelled data}
Audio tagging has attracted much research interests in recent years. For example, the tagging of the CHiME Home dataset \cite{foster2015chime}, the UrbanSound dataset \cite{salamon2014dataset} and datasets from the Detection and Classification of Acoustic Scenes and Events (DCASE) challenges in 2013 \cite{stowell2015detection}, 2016 \cite{mesaros2016tut}, 2017 \cite{mesaros2017dcase} and 2018 \cite{mesaros2018multi}. The DCASE 2018 Challenge includes acoustic scene classification \cite{mesaros2018multi}, general purpose audio tagging \cite{fonseca2018general} and bird audio detection \cite{stowell2018automatic} tasks. Mel frequency cepstral coefficients (MFCC) \cite{li2001classification, uzkent2012non, eghbal2016cp} have been widely used as features to build audio tagging systems. Other features used for audio tagging include pitch features \cite{uzkent2012non} and I-vectors \cite{eghbal2016cp}. Classifiers include such as Gaussian mixture models (GMMs) \cite{aucouturier2007bag} and support vector machines \cite{sigtia2016automatic}. Recently, neural networks have been used for audio tagging with mel spectrograms as input features. A variety of neural network methods including fully-connected neural networks \cite{kong2016deep}, convolutional neural networks (CNNs) \cite{cakir2016domestic, choi2016automatic, hershey2017cnn} and convolutional recurrent neural networks (CRNNs) \cite{choi2017convolutional, xu2017large} have been explored for audio tagging. For sound localization, an identify, locate and separate model \cite{parekh2018identify} was proposed for audio-visual object extraction in large video collections using weak supervision.

A WLD consists of a set of \textit{bags}, where each bag is a collection of instances. For a particular sound class, a positive bag contains at least one positive instance, while a negative bag contains no positive instances. We denote the $ n $-th bag in the dataset as $ B_{n}=\left \{ \mathbf{x}_{n1}, ..., \mathbf{x}_{nT_{n}} \right \} $, where $ T_{n} $ is the number of instances in the bag. An instance $ \mathbf{x}_{nt} \in \mathbb{R}^{M} $ in the bag has a dimension of $ M $. A WLD can be denoted as $ D = \left \{ B_{n}, \mathbf{y}_{n} \right \}_{n=1}^{N} $, where $ \mathbf{y}_{n} \in \{0, 1\}^{K} $ denotes the tags of bag $ B_{n} $, and $ K $ and $ N $ are the number of classes and training samples, respectively. In WLD, each bag $ B_{n} $ has associated tags but we do not know the tags of individual instances $ \mathbf{x}_{nt} $ within the bag \cite{amores2013multiple}. For example, in the AudioSet dataset, a bag consists of instances that are bottleneck features obtained by inputting a logmel to a pre-trained VGGish convolutional neural network. In the following sections, we omit the training example index $ n $ and the time index $ t $ to simplify notation. 

Previous audio tagging systems using WLD have been based on segment based methods. Each segment is called an instance and are assigned the tags inherited from the audio clip. During training, instance-level classifiers are trained on individual instances. During inference, bag-level predictions are obtained by aggregating the instance-level predictions \cite{kong2016deep}. Recently, convolutional neural networks have been applied to audio tagging \cite{choi2017convolutional}, where the log spectrogram of an audio clip is used as input to a CNN classifier without predicting the individual instances explicitly. Attention neural networks have been proposed for AudioSet tagging in \cite{kong2017audio, yu2018multi}. Later, a clip-level and segment-level model with attention supervision was proposed in \cite{chou2018learning}.

\section{Audio tagging with attention neural networks}
\subsection{Segment based methods}\label{section:bow}
R3: In segment based methods, an audio clip is split into segments and each segment is assigned the tags inherited from the audio clip. In MIL, each segment is called an instance. An instance-level classifier $ f $ is trained on the individual instances: $ f: \mathbf{x} \mapsto f(\mathbf{x}) $, where $ f(\mathbf{x}) \in [0, 1]^{K} $ predicts the presence probabilities of sound classes. The function $ f $ depends on a set of learnable parameters that can be optimised using gradient descent methods with the loss function
\begin{equation} \label{eq:bag_of_words_loss}
l(f(\mathbf{x}), \mathbf{y}) = d(f(\mathbf{x}), \mathbf{y}), 
\end{equation}
\noindent where $ \mathbf{y} \in \{0, 1\}^{K} $ are the tags of the instance $ \mathbf{x} $ and $ d(\cdot, \cdot) $ is a loss function. For instance, it could be binary cross-entropy for multi-class tagging, given by
\begin{equation} \label{eq:binary_crossentropy}
d(f(\mathbf{x}), \mathbf{y}) = - \sum_{k=1}^{K} [ y_{k} \text{log}f(\mathbf{x})_{k} + (1 - y_{k}) \text{log} (1 - f(\mathbf{x})_{k} ].
\end{equation}

\noindent In inference, the prediction of a bag is obtained by aggregating the predictions of individual instances in the bag such as by majority voting \cite{kong2016deep}. The segment based model has been applied to many tasks such as information retrieval \cite{pancoast2012bag} due to its simplicity and efficiency. However, the assumption that all instances inherit the tags of a bag is incorrect. For example, some sound events may only occur for a short time in an audio clip.

\subsection{Attention neural networks}
Attention neural networks were first proposed for natural language processing \cite{sankaran2016temporal, luong2015effective}, where the words in a sentence are attended differently for machine translation. Attention neural networks are designed to attend to important words and ignore irrelevant words. Attention models have also been applied to computer vision, such as image captioning \cite{attention-xu} and information retrieval \cite{liu2015content}. We proposed attention neural networks for audio tagging and sound event detection with WLD in \cite{kong2017audio, xu2017large}: these were ranked first in the DCASE 2017 Task 4 challenge \cite{xu2017large}. In a similar way to the segment based model, attention neural networks build an instance-level classifier $ f(\mathbf{x}) $ for individual instances $ \mathbf{x} $. In contrast to the segment based model, attention neural networks do not assume that instances in a bag have the same tags as the bag. As a result, there is no instance-level ground truth for supervised learning using (\ref{eq:bag_of_words_loss}). To solve this problem, we aggregate the instance-level predictions $ f(\mathbf{x}) $ to a bag-level prediction $ F(B) $ given by
\begin{equation} \label{eq:att_func}
F(B)_{k} = \sum_{\mathbf{x} \in B} p(\mathbf{x})_{k} f(\mathbf{x})_{k}, 
\end{equation}
where $ p(\mathbf{x})_{k} $ is a weight of $ f(\mathbf{x})_{k} $ that we refer to as an \textit{attention function}. The attention function $ p(\mathbf{x})_{k} $ should satisfy
\begin{equation} \label{eq:probability_constraint}
\sum_{\mathbf{x} \in B} p(\mathbf{x})_{k} = 1, 
\end{equation}
\noindent so that the bag-level prediction can be seen as a weighted sum of the instance-level predictions. Both the attention function $ p(\mathbf{x}) $ and the instance-level classifier $ f(\mathbf{x}) $ depend on a set of learnable parameters. The attention function $ p(\mathbf{x})_{k} $ controls how much a prediction $ f(\mathbf{x})_{k} $ should be attended. Large $ p(\mathbf{x})_{k} $ indicates that $ f(\mathbf{x})_{k} $ should be attended, while small $ p(\mathbf{x})_{k} $ indicates that $ f(\mathbf{x})_{k} $ should be ignored. To satisfy (\ref{eq:probability_constraint}), the attention function $ p(\mathbf{x})_{k} $ can be modelled as
\begin{equation} \label{eq:decision_level_normalize}
p(\mathbf{x})_{k} = v(\mathbf{x})_{k} / \sum_{\mathbf{x} \in B} v(\mathbf{x})_{k}, 
\end{equation}
\noindent where $ v(\cdot) $ can be any non-negative function to ensure that $ p(\cdot) $ is a probability. 

An extension of the attention neural network in (\ref{eq:att_func}) is the multi-level attention model \cite{yu2018multi}, where multiple attention modules are applied to utilise the hierarchical information of neural networks:
\begin{equation} \label{eq:multi_att}
F(B) = g(F_{1}(B), ..., F_{L}(B)), 
\end{equation}
\noindent where $ F_{l}(B) $ is the output of the $ l $-th attention module and $ L $ is the number of attention modules. Each $ F_{l}(B) $ can be modeled by (\ref{eq:att_func}). Then a mapping $ g $ is used to map from the predictions of $ L $ attention modules to the final prediction of a bag. The multi-level attention neural network has achieved state-of-the-art performance in AudioSet tagging.

In the next section, we show that the attention neural networks explored above can be categorised into an MIL framework. 

\section{Multiple instance learning}

Multiple instance learning (MIL) \cite{dietterich1997solving, maron1998framework} is a type of supervised learning method. Instead of receiving a set of labelled instances, the learner receives a set of labelled bags. MIL methods have many applications. For example, in \cite{dietterich1997solving}, MIL is used to predict whether new molecules are qualified to make some new drug, where molecules may have many alternative low-energy states, but only one, or some of them, are qualified to make a drug. Inspired by the MIL methods, a sound event detection system trained on WLD \cite{kumar2016audio} was proposed. General MIL methods include the expectation-maximization diversity density (EM-DD) method \cite{zhang2002dd}, support vector machine (SVM) methods \cite{andrews2003support} and neural network MIL methods \cite{zhou2002neural, wang2018revisiting}. In \cite{wang2018comparison}, several MIL pooling methods were investigated in audio tagging. Attention-based deep multiple instance learning is proposed in \cite{ilse2018attention}. 

In \cite{amores2013multiple}, MIL methods are grouped into three categories: the instance space (IS) methods, where the discriminative information is considered to lie at the instance-level; the bag space (BS) methods, where the discriminative information is considered to lie at the bag-level; and the embedded space (ES) methods, where each bag is mapped to a single feature vector that summarises the relevant information about a bag. We introduce the IS, BS and ES methods in more detail below.

\subsection{Instance space methods}\label{section:IS_paradigm}
In IS methods, an instance-level classifier $ f: \mathbf{x} \mapsto f(\mathbf{x}) $ is used to predict the tags of an instance $ \mathbf{x} $, where $ f(\mathbf{x}) \in [0, 1]^{K} $ predicts the presence probabilities of sound classes. The IS methods introduce aggregation functions \cite{amores2013multiple} to convert an instance-level classifier $ f $ to a bag-level classifier $ F: B \mapsto [0, 1]^{K} $, given by
\begin{equation} \label{eq:IS_inference_aggregation}
F(B) = \text{agg} \left ( \{f(\mathbf{x})\}_{\mathbf{x} \in B}  \right ),
\end{equation}
\noindent where $ \text{agg}(\cdot) $ is an aggregation function. The classifier $ f $ depends on a set of learnable parameters. When the IS method is trained with (\ref{eq:bag_of_words_loss}) in which each instance inherits the tags of the bag, the IS method is equivalent to the segment based model. On the other hand, the IS method can also be trained using the bag-level loss function:
\begin{equation} \label{eq:IS_loss}
l(F(B), \mathbf{y}) = d(F(B), \mathbf{y}),
\end{equation}
\noindent where $ \mathbf{y} \in \{ 0, 1 \}^{K} $ is the tag of the bag and $ d(\cdot, \cdot) $ is a loss function such as the binary cross-entropy in (\ref{eq:binary_crossentropy}).

To model the aggregation function, the standard multiple instance (SMI) assumption and collective assumption (CA) are proposed in \cite{amores2013multiple}. Under the SMI assumption, a bag-level classifier can be obtained by
\begin{equation} \label{eq:IS_max}
F(B)_{k} = \underset{\mathbf{x} \in B}{\mathrm{max}} f(\mathbf{x})_{k}, 
\end{equation}
\noindent where the subscript $ k $ denotes the $ k $-th sound class of the instance-level prediction $ f(\mathbf{x}) $ and the bag-level prediction $ F(B) $. Under the SMI assumption, for the $ k $-th sound class, only one instance with the maximum prediction probability is chosen as a positive instance. 

One problem of the SMI assumption is that a positive bag may contain more than one positive instance. In SED, some sound classes such as ``ambulance siren'' may last for several seconds and may occur in many instances. In contrast to the SMI assumption, with the CA assumption, all the instances in a bag contribute equally to the tags of the bag. The bag-level prediction can be obtained by averaging the instance-level predictions:
\begin{equation} \label{eq:IS_average}
F(B)=\frac{1}{\left | B \right |}\sum_{\mathbf{x} \in B}f(\mathbf{x}). 
\end{equation}
\noindent The symbol $ |B| $ denotes the number of instances in bag $ B $. Equation (\ref{eq:IS_average}) shows that CA is based on the assumption that all the instances in a positive bag are positive. 

\subsection{Bag space methods}\label{BS_paradigm}
Instead of building an instance-level classifier, the BS methods regard a bag $ B $ as an entirety. Building a tagging model on the bags rely on a distance function $ D(\cdot, \cdot): B \times B \mapsto \mathbb{R} $. The distance function can be, for example, the Hausdorff distance \cite{wang2000solving}:
\begin{equation} \label{eq:BS}
D(B_{1}, B_{2}) = \underset{\mathbf{x}_{1} \in B_{1}, \mathbf{x}_{2} \in B_{2}}{\textrm{min}}\left \| \mathbf{x}_{1} - \mathbf{x}_{2} \right \|. 
\end{equation}
\noindent In (\ref{eq:BS}), the distance between two bags is the minimum distance between the instances in bag $ B_{1} $ and $ B_{2} $. Then this distance function can be plugged into a standard distance-based classifier such as a k-nearest neighbour (KNN) or a support vector machine (SVM) algorithm. The computational complexity of (\ref{eq:BS}) is $ |B_{1}||B_{2}| $, which is larger than the IS and the ES methods described below.

\subsection{Embedded space methods}\label{section:ES_paradigm}
Different from the IS methods, ES methods do not classify individual instances. Instead, the ES methods define an embedding mapping from a bag to an embedding vector:
\begin{equation} \label{eq:ES_mapping}
f_{\text{emb}}: B \mapsto \mathbf{h}. 
\end{equation}
\noindent Then the tags of a bag is obtained by applying a function $ g $ on the embedding vector:
\begin{equation} \label{eq:ES_prediction}
F(B) = g(\mathbf{h}). 
\end{equation}
\noindent The embedding mapping $ f_{\text{emb}} $ can be modelled in many ways. For example, by averaging the instances in a bag, as in the simple MI method in \cite{dong2006comparison}:
\begin{equation} \label{eq:ES_average}
\mathbf{h}=\frac{1}{\left | B \right |}\sum_{\mathbf{x} \in B}\mathbf{x}. 
\end{equation}
\noindent Alternatively, the mapping can be obtained in terms of the max-min operations on the instances \cite{gartner2002multi}:
\begin{equation} \label{eq:ES_max}
\begin{split}
\begin{cases}
\mathbf{h}=(a_{1}, ..., a_{M}, b_{1}, ..., b_{M}), \\ 
a_{m}=\underset{\mathbf{x} \in B}{\mathrm{max}} (x_{m}), \\
b_{m}=\underset{\mathbf{x} \in B}{\mathrm{min}} (x_{m}), \\
\end{cases}
\end{split}
\end{equation}
\noindent where $ x_{m} $ is the $ m $-th dimension of $ \mathbf{x} $. Equation (\ref{eq:ES_max}) shows that only one instance with the maximum or the minimum value is chosen for each dimension, while other instances have no contribution to the embedding vector $ \mathbf{h} $. The ES methods summarise a bag containing an arbitrary number of instances with a vector of fixed size. Similar methods have been proposed in natural language processing to summarise sentences with a variable number of words \cite{bahdanau2014neural}. 

\section{Attention neural networks under MIL}\label{section:revision_attention_neural_network}

In this section, we show that the previously proposed attention neural networks \cite{kong2017audio, yu2018multi} belong to MIL frameworks, especially the IS methods. We refer to these attention neural networks as decision-level attention neural networks, because the prediction of a bag is obtained by aggregating the predictions of instances (see (\ref{eq:IS_inference_aggregation})). We then propose feature-level attention neural networks inspired by the ES methods with attention in the hidden layers.

\subsection{Decision-level attention neural networks}\label{section:decision_level_att}

The IS methods predict the tags of a bag by aggregating the predictions of individual instances in the bag described in (\ref{eq:IS_inference_aggregation}). Section \ref{section:IS_paradigm} shows that conventional IS methods are based on either the SMI assumption (see (\ref{eq:IS_max})) or the CA (see (\ref{eq:IS_average})). The problem of the SMI assumption is that only one instance in a bag is considered to be positive for a sound class while other instances are not considered. The SMI assumption is not appropriate for bags with more than one positive instance for a sound class. On the other hand, CA assumes that all instances in a positive bag are positive. CA is not appropriate for sound events that only last for a short time. To address the problems of the SMI and CA methods, a decision-level attention neural network based on the IS methods in (\ref{eq:IS_inference_aggregation}) is proposed to learn an attention function to weight the predictions of instances in a bag, so that
\begin{equation} \label{eq:decision_level_agg}
\begin{split}
F(B)_{k} & = \text{agg}(\{f(\mathbf{x})_{k}\}_{\mathbf{x} \in B}) \\
& = \sum_{\mathbf{x} \in B} p(\mathbf{x})_{k} f(\mathbf{x})_{k},
\end{split}
\end{equation}
\noindent where $ p(\mathbf{x}) $ is an attention function modelled by (\ref{eq:decision_level_normalize}). We refer to (\ref{eq:decision_level_agg}) as a decision-level attention neural network because the attention function $ p(\mathbf{x}) $ is multiplied with the predictions of the instances $ f(\mathbf{x}) $ to obtain the bag-level prediction. The attention function $ p(\mathbf{x}) $ controls how much the prediction of an instance $ f(\mathbf{x}) $ should be attended or ignored. Equation (\ref{eq:decision_level_agg}) can be seen as a general case of the SMI and CA assumptions. When one instance $ \mathbf{x} $ in a bag has a value of $ p(\mathbf{x})=1 $ the other instances have values of $ p(\mathbf{x})=0 $, then (\ref{eq:decision_level_agg}) is equivalent to the SMI assumption in (\ref{eq:IS_max}). When $ p(\mathbf{x})=\frac{1}{|B|} $ for all instances in a bag, (\ref{eq:decision_level_agg}) is equivalent to CA. 

\begin{figure}[t]
  \centering
  \centerline{\includegraphics[width=\columnwidth]{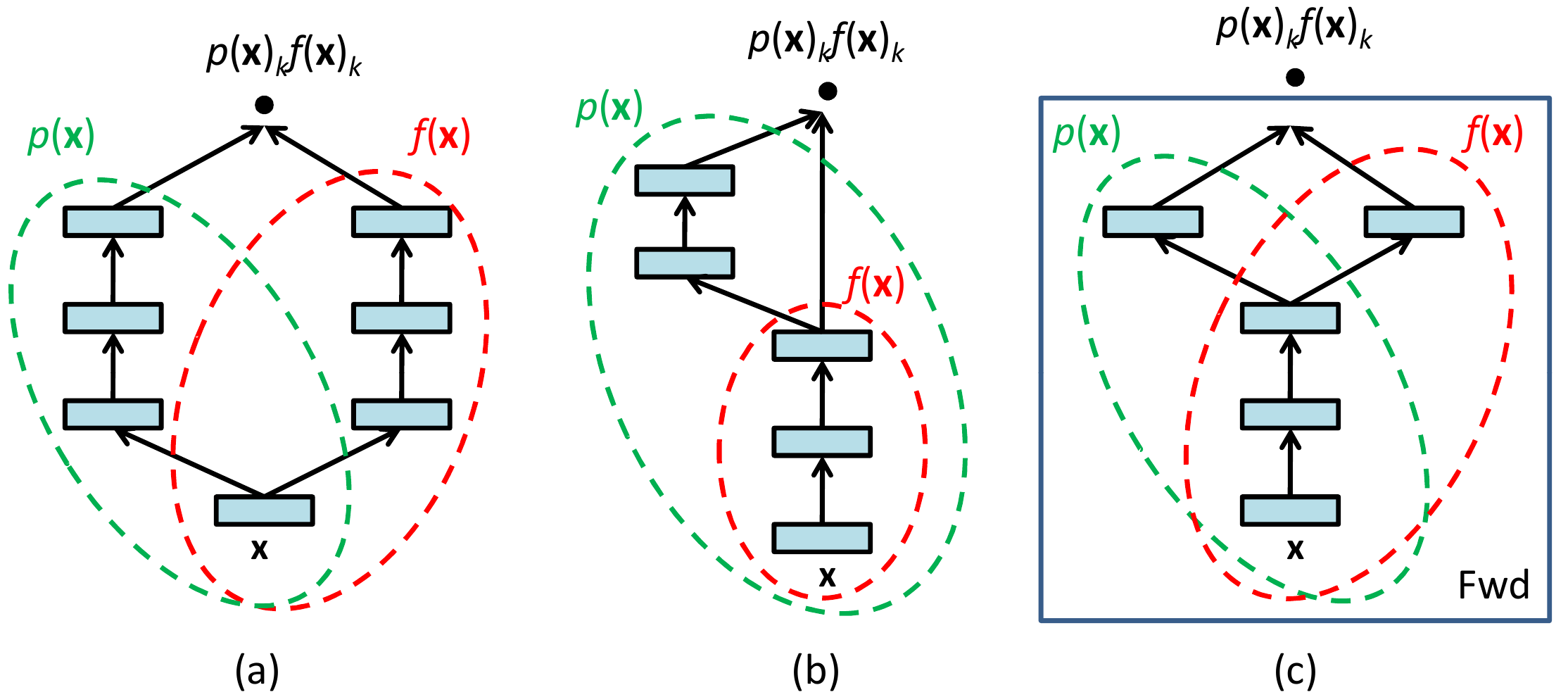}}
  \caption{(a) Joint detection and classification (JDC) model; (b) Self attention neural network in \cite{lin2017structured}; (c) Proposed attention neural network \cite{kong2017audio}. The blue outlined block in (c) is called a forward (FWD) block.}
  \label{fig:jdc_att}
\end{figure}

Fig. \ref{fig:jdc_att} shows different ways to model the attention neural network in (\ref{eq:decision_level_agg}). For example, Fig. \ref{fig:jdc_att}(a) shows the joint detection and classification (JDC) model \cite{kong2017joint} with attention function $ p $ and the classifier $ f $ modelled by separate branches. Fig. \ref{fig:jdc_att}(b) shows the self attention neural network \cite{lin2017structured} proposed in natural language processing. Fig. \ref{fig:jdc_att}(c) shows the JDC improved by using shared layers for the attention function $ p $ and the classifier $ f $ before they separate in the penultimate layer \cite{kong2017audio}. 

In the attention neural networks, both $ p $ and $ f $ depend on a set of learnable parameters which can be optimised with gradient descent methods using the loss function in (\ref{eq:IS_loss}). For the proposed model in Fig. \ref{fig:jdc_att}(c), the attention function $ p $ and the classifier $ f $ share the low-level layers. We denote the output of the layer before they separate as $ \mathbf{x}' $. The mapping from $ \mathbf{x} $ to $ \mathbf{x}' $ can be modelled by fully-connected layers, for example.
\begin{equation} \label{eq:att_fc_func}
\mathbf{x}' = f_{\text{FC}}(\mathbf{x}). 
\end{equation}
The classifier $ f $ can be modelled by
\begin{equation} \label{eq:decision_level_att_f}
f(\mathbf{x}) = \sigma(\mathbf{W}_{1} \mathbf{x}' + \mathbf{b}_{1}),
\end{equation}
\noindent where $ \sigma(x) = 1 / (1 + e^{-x}) $ is the sigmoid function. The attention function $ p $ can be modelled by
\begin{equation} \label{eq:decision_level_att_p}
\begin{cases}
v(\mathbf{x}')_{k} = \phi_{1}(\mathbf{U}_{1}\mathbf{x}' + \mathbf{c}_{1}), \\
p(\mathbf{x})_{k} = v(\mathbf{x}')_{k} / \sum_{\mathbf{x} \in B} v(\mathbf{x}')_{k}, \\
\end{cases}
\end{equation}
\noindent where $ \phi_{1} $ can be any non-negative function to ensure $ p(\mathbf{x})_{k} $ is a probability. 

\begin{figure}[t]
  \centering
  \centerline{\includegraphics[width=\columnwidth]{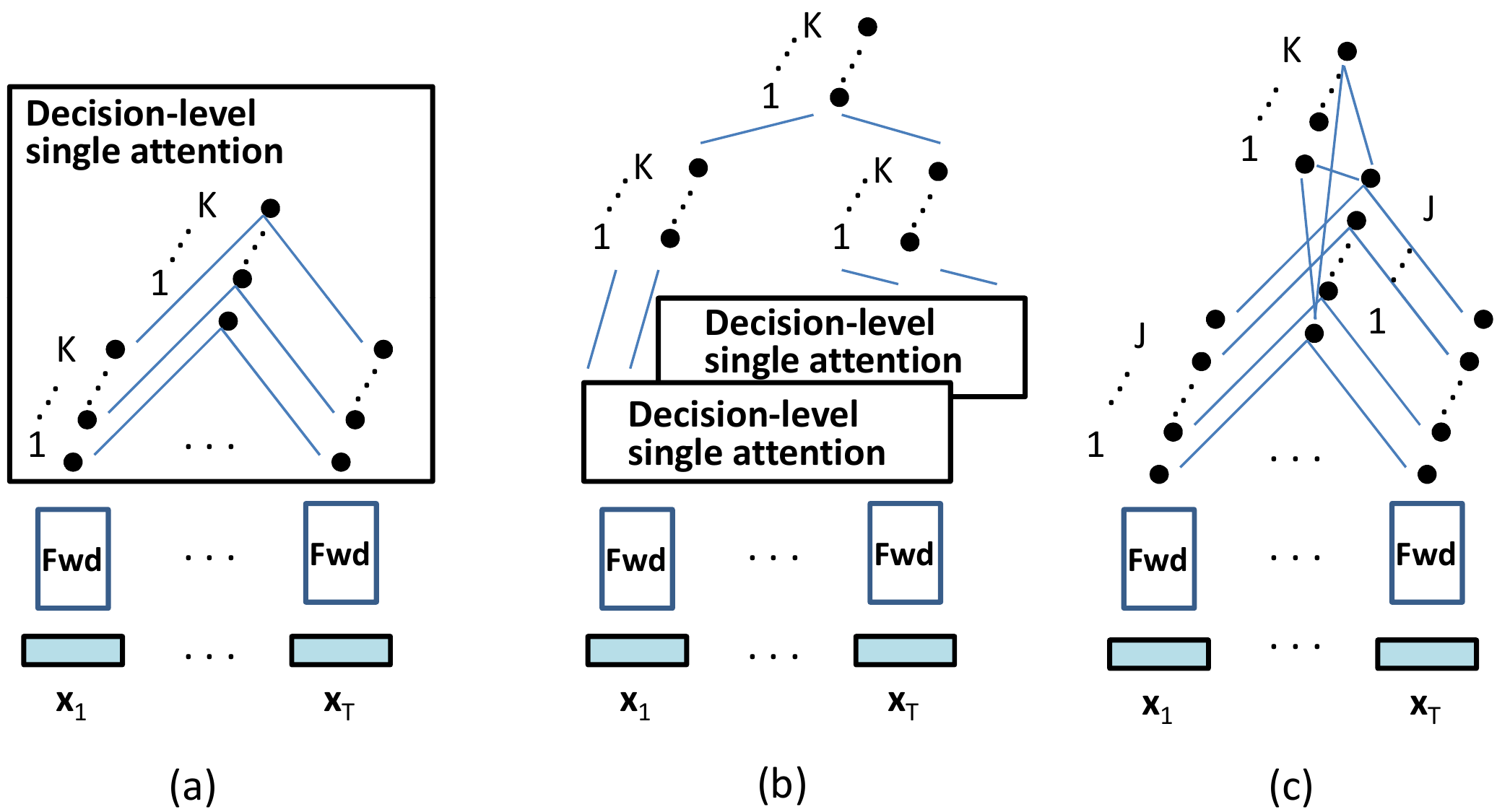}}
  \caption{(a) Decision-level single attention neural network \cite{kong2017audio}; (b) Decision-level multiple attention neural network \cite{yu2018multi}; (c) Feature-level attention neural network (proposed). }
  \label{fig:decision_feature_att}
\end{figure}

\subsection{Feature-level attention neural network}\label{section:feature_level_att}
The limitation of the decision-level attention neural networks is that the attention function $ p(\mathbf{x}) $ is only applied to the prediction of the instances $ f(\mathbf{x}) $, as shown in (\ref{eq:decision_level_agg}). In this section, we propose to apply attention to the hidden layers of a neural network. This is inspired by the ES methods in (\ref{eq:ES_mapping}), where a bag $ B $ is mapped to a fixed-size vector $ \mathbf{h} $ before being classified. We model (\ref{eq:ES_mapping}) with attention aggregation:
\begin{equation} \label{eq:feature_level_att_1}
h_{j} = \sum_{\mathbf{x} \in B} q(\mathbf{x})_{j} u(\mathbf{x})_{j},
\end{equation}
\noindent where both $ q(\mathbf{x}) \in [0, 1]^{J} $ and $ u(\mathbf{x}) \in \mathbb{R}^{J} $ have a dimension of $ J $. The embedded vector $ \mathbf{h} \in \mathbb{R}^{J} $ summarises the information of a bag. Then the tags of a bag $ B $ can be obtained by classifying the embedding vector: 
\begin{equation} \label{eq:feature_level_att_2}
F(B) = f(\mathbf{h}).
\end{equation}
The probability $ q(\mathbf{x})_{j} $ in (\ref{eq:feature_level_att_1}) is the attention function of $ u(\mathbf{x})_{j} $ and should satisfy
\begin{equation} \label{eq:feature_level_constraint}
\sum_{\mathbf{x} \in B} q(\mathbf{x})_{j} = 1.
\end{equation}
We model $ u(\mathbf{x}) $ with
\begin{equation} \label{eq:feature_level_att_f}
u(\mathbf{x}) = \psi(\mathbf{W}_{2} \mathbf{x}' + \mathbf{b}_{1}),
\end{equation}
\noindent where $ \psi $ can be any linear or non-linear function to increase the representation ability of the model. The attention function $ q $ can be modelled by
\begin{equation} \label{eq:feature_level_att_q}
\begin{cases}
w(\mathbf{x}')_{j} = \phi_{2}(\mathbf{U}_{2}\mathbf{x}' + \mathbf{c}_{2}), \\
q(\mathbf{x})_{j} = w(\mathbf{x}')_{j} / \sum_{\mathbf{x} \in B} w(\mathbf{x}')_{j}, \\
\end{cases}
\end{equation}
\noindent where $ w(\mathbf{x})_{j} $ can be any non-negative function to ensure $ q(\mathbf{x})_{j} $ is a probability. 

Fig. \ref{fig:decision_feature_att} shows the decision-level single attention \cite{kong2017audio}, decision-level multiple attention \cite{yu2018multi} and the proposed feature-level attention neural network. The forward (Fwd) block in Fig. \ref{fig:decision_feature_att} is the same as the block in Fig. \ref{fig:jdc_att}(c). The difference between the feature-level attention function $ q(\mathbf{x}) $ and the decision-level attention function $ p(\mathbf{x}) $ is that the dimension of $ q(\mathbf{x}) $ can be any value, while the dimension of $ p(\mathbf{x}) $ is fixed to be the number of sound classes $ K $. Therefore, the capacity of the decision-level attention neural networks is limited. With an increase in the dimension of $ q(\mathbf{x}) $, the capacity of feature-level attention neural networks is increased. The decision-level attention function attends to the predictions of instances, while the feature-level attention function attends to the features, so it is equivalent to feature selection. The multi-level attention model \cite{yu2018multi} in (\ref{eq:multi_att}) can be seen as a special case of the feature-level attention model, with embedding vector $ \mathbf{h} = (F_{1}(B), ..., F_{L}(B)) $. The superior performance of the multi-level attention model shows that the feature-level attention neural networks have the potential to perform better than the decision-level attention neural networks. 

\subsection{Modeling the attention function with different non-linearity}\label{section:exp_different_att_func}

We adopt Fig. \ref{fig:jdc_att}(c) as the backbone of our attention neural networks. The attention function $ p $ and $ q $ for the decision-level and feature-level attention neural networks are obtained via non-negative functions $ \phi_{1} $ and $ \phi_{2} $, respectively. The $ \phi_{1} $ and $ \phi_{2} $ appearing in the summation term of the denominator of (\ref{eq:decision_level_att_p}) and (\ref{eq:feature_level_att_q}) may affect the optimisation of the attention neural networks. We investigate modelling $ \phi_{2} $ in the feature-level attention neural networks with different non-negative functions, including ReLU \cite{nair2010rectified}, exponential, sigmoid, softmax and network-in-network (NIN) \cite{lin2013network}. We omit the evaluation of $ \phi_{1} $, as the feature-level attention neural networks outperform the decision-level attention neural networks. The ReLU function is defined as \cite{nair2010rectified}
 \begin{equation} \label{eq:relu}
 \phi(z) = \text{max}(z, 0). 
\end{equation}
The exponential function is defined as
 \begin{equation} \label{eq:exponential}
 \phi(z) = e^{z}.
\end{equation}
The sigmoid function is defined as
\begin{equation} \label{eq:sigmoid}
\phi(z) = \frac{1}{1 + e^{-z}}. 
\end{equation}
For a vector $\mathbf{z}$, the softmax function is defined as
\begin{equation} \label{eq:softmax}
\phi(z_{j}) = \frac{e^{z_{j}}}{\sum_{k}e^{z_{k}}}.
\end{equation}
The network-in-network function \cite{lin2013network} is defined as
\begin{equation} \label{eq:nin}
\phi(\mathbf{z}) = \sigma(\mathbf{H}_{2}\psi(\mathbf{H}_{1}\mathbf{z}+\mathbf{d}_{1})+\mathbf{d}_{2}),
\end{equation}
\noindent where $ \mathbf{H}_{1} $, $ \mathbf{H}_{2} $ are transformation matrices, $ \mathbf{d}_{1} $ and $ \mathbf{d}_{2} $ are biases, $\psi$ is ReLU nonlinearity and $ \sigma $ is the sigmoid function.

\begin{figure}[t]
  \centering
  \centerline{\includegraphics[width=\columnwidth]{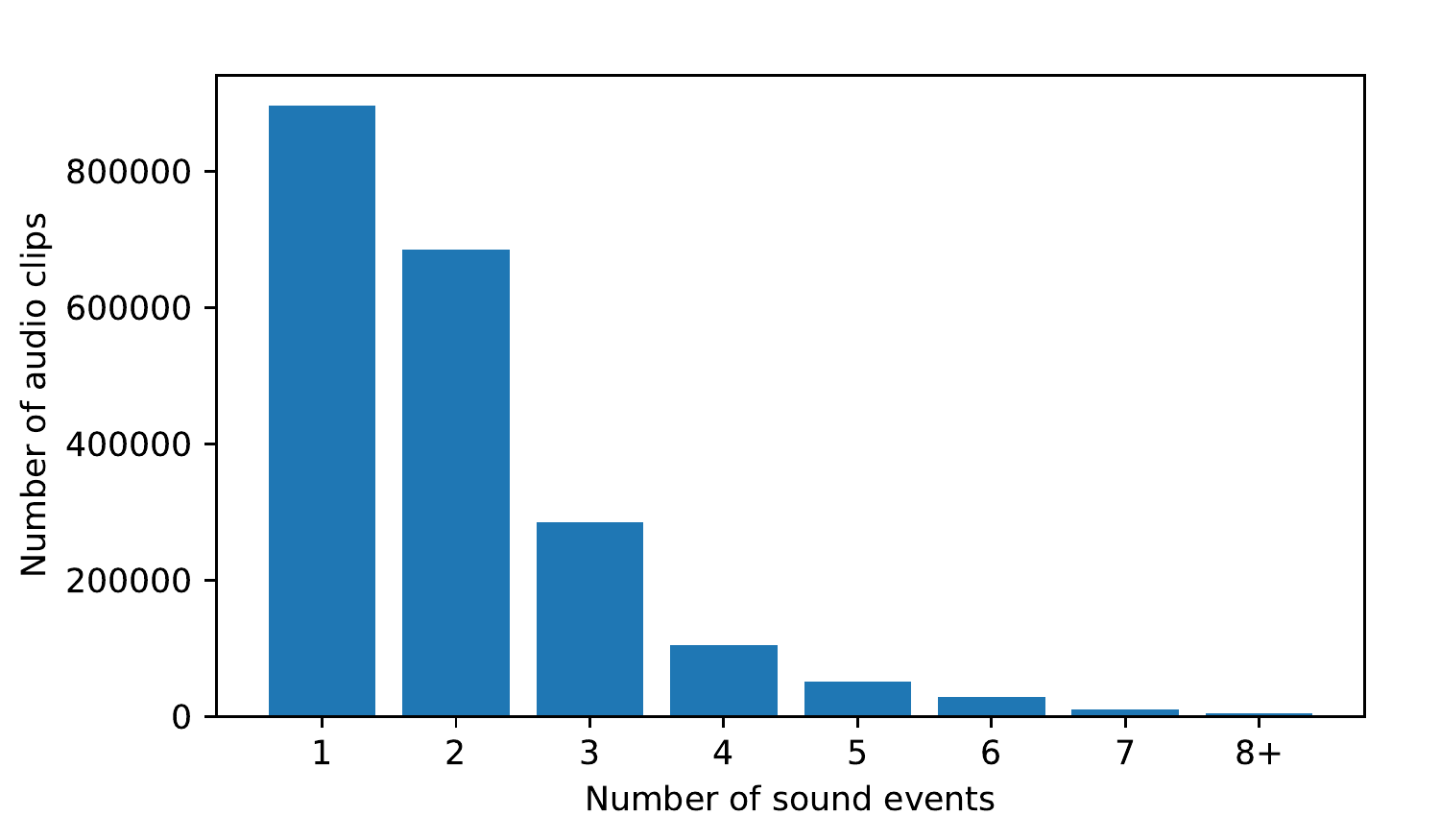}}
  \caption{Distribution of the number of sound classes in an audio clip. }
  \label{fig:classes_per_clip}
\end{figure}

\begin{figure}[t]
  \centering
  \centerline{\includegraphics[width=\columnwidth]{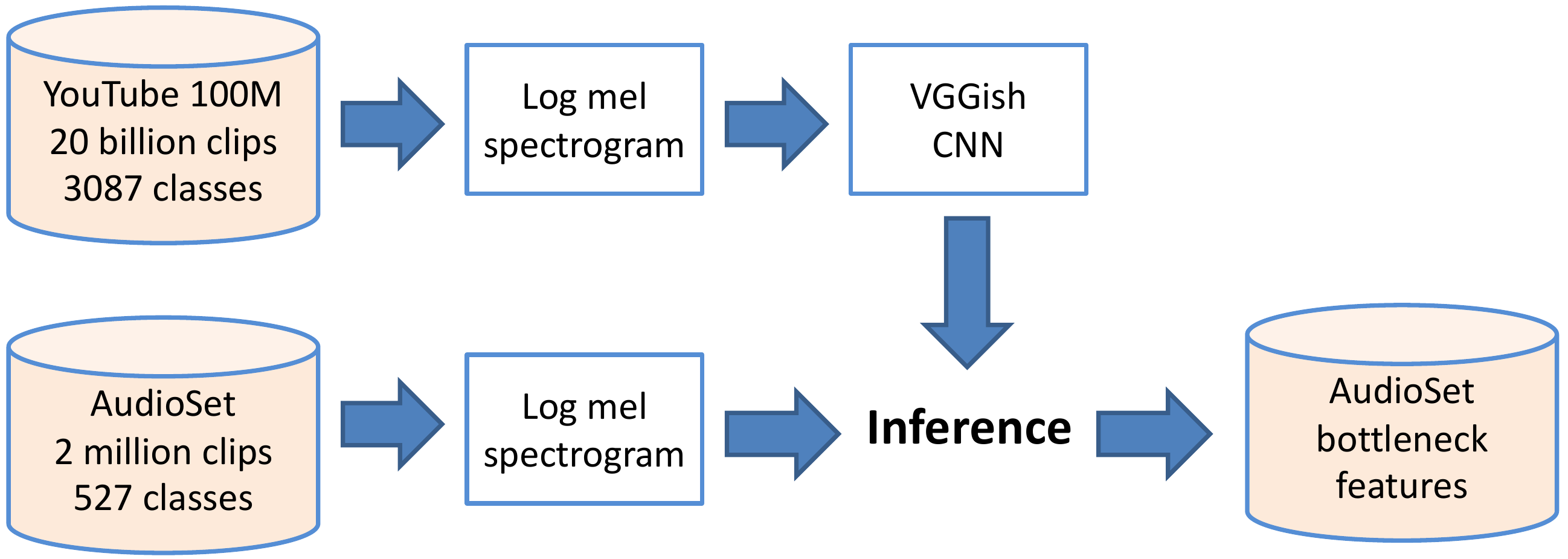}}
  \caption{A VGGish CNN model is trained on the YouTube 100M dataset. Audio clips from AudioSet are given as input to the trained VGGish CNN to extract the bottleneck features, which are released by AudioSet. }
  \label{fig:feature_extraction}
\end{figure}

\begin{figure*}[t]
  \centering
  \centerline{\includegraphics[width=\textwidth]{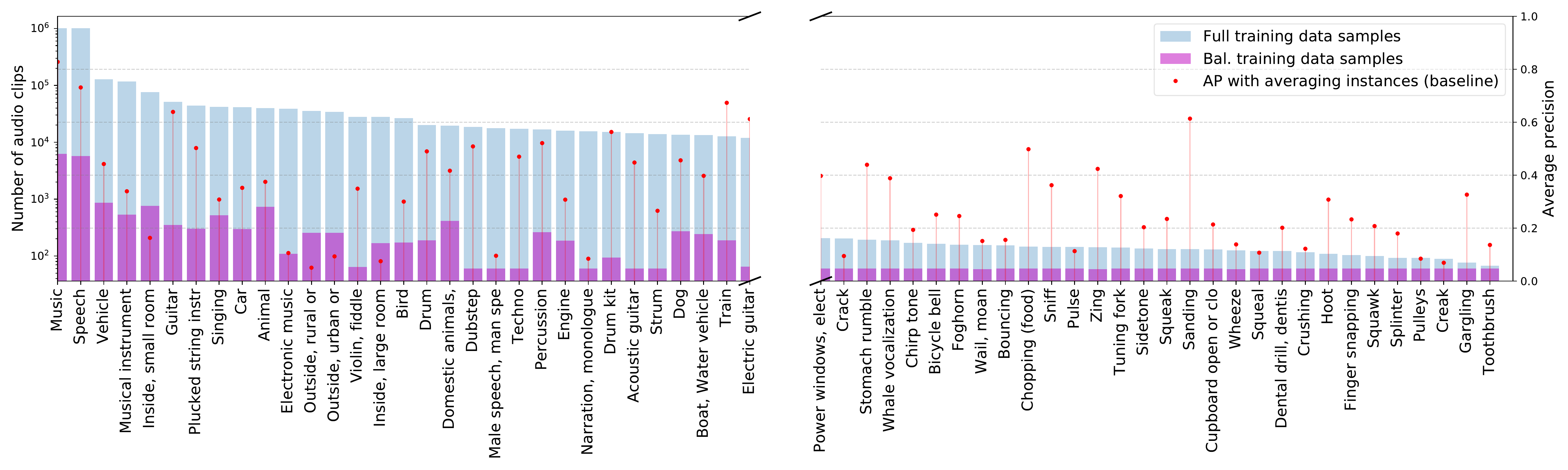}}
  \caption{AudioSet statistics. Upper bars: the number of audio clips of a specific sound class sorted in descending order plotted in log scale with respect to the sound classes. Red stems: average precision (AP) of sound classes with the feature-level attention model. }
  \label{fig:data_distribution}
\end{figure*}

\section{Experiments}
\subsection{Dataset}
We evaluate the proposed attention neural networks on AudioSet \cite{audioset}, which consists of 2,084,320 10-second audio clips extracted from YouTube video with a hierarchical ontology of 527 classes in the released version (v1). We released both Keras and PyTorch implementations of our code online\footnote{https://github.com/qiuqiangkong/audioset\textunderscore classification}. AudioSet consists of a variety of sounds. AudioSet is multi-labelled, such that each audio clip may contain more than one sound class. Fig. \ref{fig:classes_per_clip} shows the statistics of the number of sound classes in the audio clips. All audio clips contain at least one label. Out of over 2,084,320 audio clips, there are 896,045 audio clips containing one sound class, followed by around 684,166 audio clips containing two sound classes. Only 4,661 audio clips have more than 7 labels. 

Instead of providing raw audio waveforms, AudioSet provides bottleneck features of audio clips. The bottleneck features are extracted from the bottleneck layer of a VGGish CNN, pre-trained on 70 million audio clips from the YouTube100M dataset \cite{hershey2017cnn}. The VGGish CNN consists of 6 convolutional layers with kernel size of $ 3 \times 3 $ and 2 fully  layers. To begin with, the 70 million training audio clips are segmented to non-overlapping 960 ms segments. Each segment inherits all tags of its parent video. Then short-time Fourier transform (STFT) is applied on each 960 ms segment with a window size of 25 ms and a hop size of 10 ms to obtain a spectrogram. Then a mel filter bank with 64 frequency bins is applied on the spectrograms followed by a logarithmic operation to obtain log mel spectrograms. Each log mel spectrogram of a segment has a shape of $ 96 \times 64 $, representing the time steps and the number of mel frequency bins. A VGGish CNN is trained on these log mel spectrograms with the 3087 most frequent labels. After training, the VGGish CNN is used as a feature extractor. By inputting an audio clip to the VGGish CNN, the outputs of the bottleneck layer are used as bottleneck features of the audio clip. The framework of AudioSet feature extraction is shown in Fig. \ref{fig:feature_extraction}. 

\begin{table}[t]
\centering
\caption{Baseline results of segment based method, IS and ES methods}
\label{table:baseline}
\begin{tabular}{*{4}{c}}
 \toprule
 & mAP & AUC & d-prime \\
 \midrule
 Random guess & 0.005 & 0.500 & 0.000 \\
 Google baseline \cite{audioset} & 0.314 & 0.959 & 2.452 \\
 \midrule
 Segment based \cite{kong2016deep} & 0.293 & 0.960 & 2.483 \\
 \midrule
 (IS) SMI assumption \cite{kumar2016audio} & 0.292 & 0.960 & 2.471 \\
 (IS) Collective assumption & 0.300 & 0.964 & 2.536 \\
 \midrule
 (ES) Average instances \cite{dong2006comparison} & \textbf{0.317} & \textbf{0.963} & \textbf{2.529} \\
 (ES) Max instance & 0.284 & 0.958 & 2.443 \\
 (ES) Min instance & 0.281 & 0.956 & 2.413 \\
 (ES) Max-min instance \cite{gartner2002multi} & 0.306 & 0.962 & 2.505 \\
 \bottomrule
\end{tabular}
\end{table}

\subsection{Evaluation criterion}
We first introduce basic statistics \cite{mesaros2016metrics}: true positive (TP), where both the reference and the system prediction indicate an event to be active; false negative (FN), where the reference indicates an event is active but the system prediction indicates an event is inactive; false positive (FP), where the system prediction indicates an event is active but the reference indicates it is inactive; true negative (TN), where both the reference and the system prediction indicate an event is inactive. Precision (P) and recall (R) are defined as in \cite{mesaros2016metrics}:

\begin{equation} \label{eq:prec_recall}
\text{P}=\frac{\text{TP}}{\text{TP}+\text{FP}}, \qquad \text{R}=\frac{\text{TP}}{\text{TP}+\text{FN}}.
\end{equation}
In addition, the false positive rate is defined as \cite{mesaros2016metrics}
\begin{equation} \label{eq:fpr}
\text{FPR} = \frac{\text{FP}}{\text{FP} + \text{TN}}. 
\end{equation}
 Following \cite{audioset}, we adopt mean average precision (mAP), area under the curve (AUC) and d-prime as evaluation metrics. Average precision (AP) \cite{audioset} is defined as the area under the recall-precision curve of a specific class. The mean average precision (mAP) is the average value of AP over all classes. As AP is regardless of TN, AUC is used as a complementary metric. AUC is the area under the receiver operating characteristic (ROC) created by plotting the recall against the false positive rate (FPR) at various threshold settings for a specific class. We use mAUC to denote the average value of AUC over all classes. D-prime is a statistic used in signal detection theory that provides separation between signal and noise distributions. D-prime is obtained via a transformation of AUC and has a better dynamic range than AUC when AUC is larger than 0.9. A higher mAP, AUC and d-prime indicates a better performance. D-prime can be calculated by \cite{audioset}:
 \begin{equation} \label{eq:auc}
\text{d-prime}=\sqrt{2}F_{x}^{-1}(\text{AUC}), 
\end{equation}
where $F_{x}^{-1}$ is an inverse of the cumulative distribution function defined by
\begin{equation}
    F_{x}(x) = \int_{-\infty}^{x}\frac{1}{\sqrt{2\pi}}e^{\frac{-({x-\mu})^2}{2}}dx  \tag{8} \label{eq:8}. 
\end{equation}

\begin{table}[t]
\centering
\caption{Results of ES average instances method with different balancing strategy. }
\label{table:balancing}
\begin{tabular}{*{4}{c}}
 \toprule
 & mAP & AUC & d-prime \\
 \midrule
 Balanced data & 0.274 & 0.949 & 2.316 \\
 Full data (no bal. training) & 0.268 & 0.950 & 2.331 \\
 Full data (bal. training) & \textbf{0.317} & \textbf{0.963} & \textbf{2.529} \\
 \bottomrule
\end{tabular}
\end{table}

\subsection{Baseline system}\label{section:experiment_baseline}
We build baseline systems with segment based method, IS and ES models without the attention mechanism described in Section \ref{section:bow}, \ref{section:IS_paradigm} and \ref{section:ES_paradigm}, respectively. In the segment based model, a classifier is trained on individual instances, where each instance inherits the tags of a bag. A three-layer fully-connected neural network with 1024 hidden units and ReLU \cite{nair2010rectified} non-linearity is applied. Dropout \cite{srivastava2014dropout} with a rate of 0.5 is used to prevent overfitting. The loss function for training is given in (\ref{eq:bag_of_words_loss}). In inference, the prediction is obtained by averaging the prediction of individual instances. The IS models have the same structure as the segment based model. Different from the segment based model, the instance-level predictions by the IS models are aggregated to a bag-level prediction by either the SMI assumption in (\ref{eq:IS_max}) or CA in (\ref{eq:IS_average}). The loss function is calculated from (\ref{eq:IS_loss}). The ES method aggregates the instances of a bag to an embedded vector before tagging. The embedding function can be the averaging mapping in (\ref{eq:ES_average}) or max-min vector mapping in (\ref{eq:ES_max}). Then the embedded vector is input to a neural network in the same way as the segment based model. The loss function is calculated from (\ref{eq:IS_loss}). We adopt the Adam optimiser \cite{kingma2014adam} with a learning rate of 0.001 in training. The mini-batch size is set to 500. The networks are trained for a total number of 50,000 iterations. We average the predictions of 9 models from 10,000 to 50,000 iterations as the final prediction to ensemble and stabilise the result, which can reduce the prediction randomness caused by the model. 

Table \ref{table:baseline} shows the tagging result of segment based method, IS and ES baseline methods. The first row shows that the random guess achieves an mAP of 0.005, an AUC of 0.500 and a d-prime of 0. The segment based model achieves an mAP of 0.293, slightly better than the IS methods with the CA and SMI assumption, with mAP of 0.300 and 0.292, respectively. The sixth to the ninth rows show that both the ES methods with averaging and the max-min instances perform better than the segment based model and IS methods. Averaging the instances performs the best in the ES methods with an mAP of 0.317, an AUC of 0.963 and a d-prime of 2.529.

\subsection{Data balancing}
AudioSet is highly imbalanced, as some sound classes such as speech and music are more frequent than others. The upper bars in Fig. \ref{fig:data_distribution} show the number of audio clips per class sorted in descending order (in log scale). The data has a long tail distribution. Music and speech appear in almost 1 million audio clips while some sounds such as gargling and toothbrush only appear in hundreds of audio clips. AudioSet provides a balanced subset consisting of 22,160 audio clips. The lower bars in Fig. \ref{fig:data_distribution} show the number of audio clips per class of the balanced subset. When training a neural network, data is loaded in mini batches. We found that without a balancing strategy, the classes with fewer samples are less likely to be selected in training. Several balancing strategies have been investigated in image classification such as balancing the frequent and infrequent classes \cite{shen2016relay}. In this paper, we follow the mini-batch balancing strategy \cite{kong2017audio} for AudioSet tagging, where each mini-batch is balanced to have approximately the same number of samples in training the neural network. 

\begin{figure*}[t]
  \centering
  \centerline{\includegraphics[width=\textwidth]{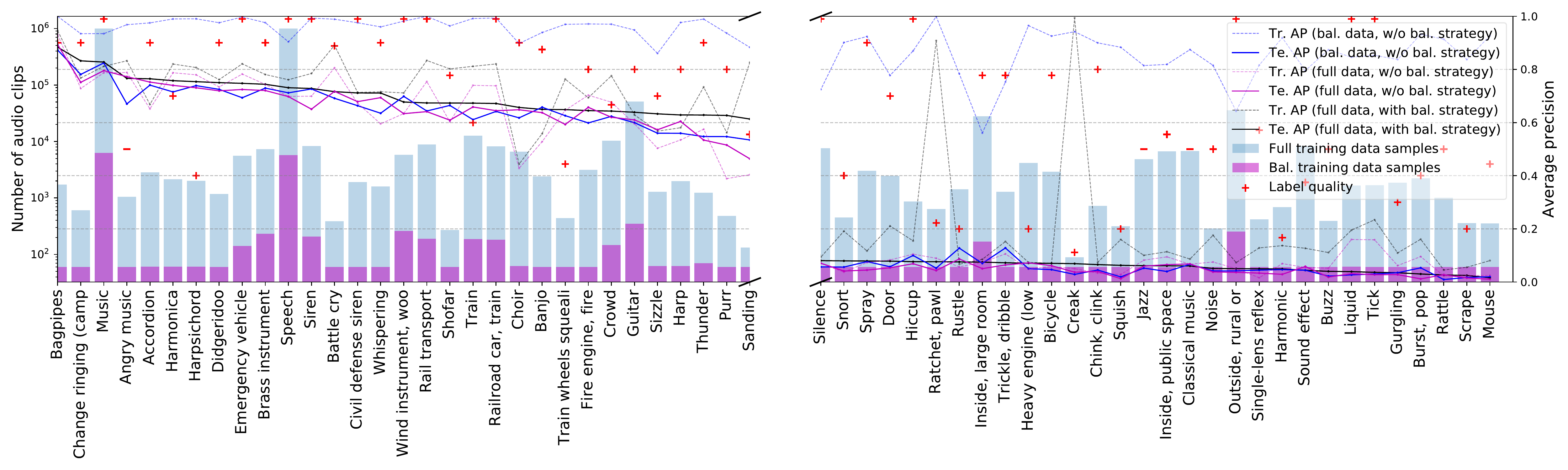}}
  \caption{Class-wise AP of sound events using the IS average instances model trained with different balancing strategy. Abbreviations: Tr.: Training; Te.: Testing; bal.: training with balanced subset; full: trained with full dataset; w/o: without mini-batch data balancing; w.: with mini-batch data balancing. }
  \label{fig:balance}
\end{figure*}

\begin{figure*}[t]
  \centering
  \centerline{\includegraphics[width=\textwidth]{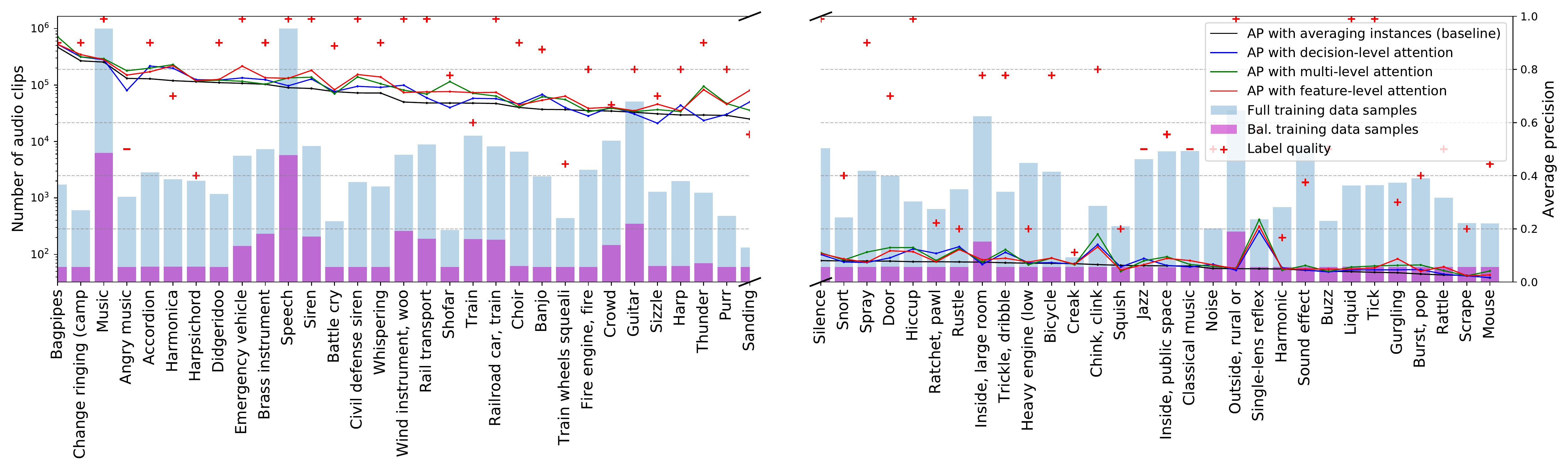}}
  \caption{Class-wise AP of sound events predicted using different models.  }
  \label{fig:aggregation}
\end{figure*}

We first investigate the performance of training on the balanced subset only and training on the full data. We adopt the best baseline model; that is, the ES average instances model in Section \ref{section:ES_paradigm}. Table \ref{table:balancing} shows that the model trained with only the balanced subset achieves an mAP of 0.274. The model trained with the full dataset without balancing achieves an mAP of 0.268. The model trained with the balancing strategy achieves an mAP of 0.317. Fig. \ref{fig:balance} shows the class-wise AP. The dashed and solid curves show the training and testing AP, respectively. In addition, Fig. \ref{fig:balance} shows that the AP is not always positive related to the number of training samples. For example, when using full data for training, ``bagpipes'' has 1,715 audio clips but achieves an mAP of 0.884, while ``outside'' has 34,117 audio clips but only achieves an AP of 0.093. We discover that for a majority of sound classes, the improvement of AP is small compared when using the full dataset rather than the balanced subset. For example, there are 60 and 1,715 ``bagpipes'' audio clips in the balanced subset and the full dataset, respectively. Their APs are 0.873 and 0.884, respectively, indicating that collecting more data for ``bagpipes'' does not substantially improve its tagging result.

To investigate how AP is related to the number of training samples, we calculate their Pearson correlation efficient (PCC)\footnote{Given a pair of random variables $ X $ and $ Y $, the PCC is calculated as $ \frac{\text{cov(X, Y)}}{\sigma_{X}\sigma_{Y}} $, where $ \text{cov}(\cdot, \cdot) $ is the covariance of two variables and $ \sigma $ is the standard deviation of the random variables. }. PCC is a number between -1 and +1. The PCC of -1, 0, +1 indicate negative correlation, no correlation and positive correlation, respectively. The null hypothesis is that the correlation of the pair of random variables is 0. The p-value indicates the confidence when the null hypothesis is satisfied. If the p-value is lower than the conventional 0.05 the PCC is called statistically significant. Table \ref{table:PCC} shows that AP and the number of training samples have a correlation with a PCC of 0.169 and the p-value is $ 9.35 \times 10 ^{-5} $, indicating that AP is only weakly positively related with the number of training samples. 

\begin{table}[t!]
\centering
\caption{Correlation of mAP with training samples and labels quality of sound classes. }
\label{table:PCC}
\begin{tabular}{*{3}{c}}
 \toprule
 & PCC & p-value \\
 \midrule
 Training examples & 0.169 & 9.35 $ \times 10^{-5} $ \\
 Labels quality & 0.230 & 7 $\times 10^{-7} $ \\
 \bottomrule
\end{tabular}
\end{table}

\subsection{Noisy labels}
AudioSet contains noisy tags \cite{audioset}. That is, some tags for training may be incorrect. There are three major reasons leading to the noisy tags in AudioSet shown in \cite{audioset}: 1) confusing labels, where some sound classes are easily confused with others; 2) human error, where the labeling procedure may be flawed; 3) faint/non-salient sounds, where some sound are faint to recognise in an audio clip. Sound classes with a high label confidence include ``christmas music'' and ``accordion''. Sound classes with a low label confidence include ``boiling'' and ``bicycle''. To investigate how accurate are the ground truth tags, The authors of AudioSet conducted an internal quality assessment task where experts checked 10 random segments for most of the classes. The quality is a value between 0 and 1 measured by the percentage of correctly labelled audio clips verified by human. The quality of labels is shown in Fig. \ref{fig:balance} with red plus symbols. Hyphen symbols are plotted for the classes that have not been evaluated. We discover that AP is not always correlated positively with the quality of labels. For example, our model achieves an AP of 0.754 in recognizing ``harpsichord'', while the human label quality is 0.4. On the other hand, humans achieve a label quality of 1.0 in ``hiccup'', but the AP of our model is 0.076. Table \ref{table:PCC} shows that AP and the quality of labels have a weak PCC of 0.230, indicating AP is only weakly correlated with the quality of labels.

\subsection{Attention neural networks}
We evaluate the decision-level and the feature-level attention neural networks in this subsection. We adopt the architecture in Fig. \ref{fig:jdc_att}(c) as our model. The output $ \mathbf{x}' $ of the layer before the attention function is obtained by (\ref{eq:att_fc_func}). Then the decision-level and feature-level attention neural networks are modelled by (\ref{eq:decision_level_agg}) and (\ref{eq:feature_level_att_1}), respectively. The first row of Table \ref{table:attention} shows that the ES method with averaged instances achieves an mAP of 0.317. The second and third rows show that the JDC model in Fig. \ref{fig:jdc_att}(a) and the self-attention model in Fig. \ref{fig:jdc_att}(b) achieve an mAP of 0.337 and 0.324, respectively. The fourth and fifth row show that the decision-level attention neural network achieves an mAP of 0.337. The decision-level multiple attention neural network further improves this result to an mAP of 0.357. 

\begin{figure*}[t!]
  \centering
  \centerline{\includegraphics[width=\textwidth]{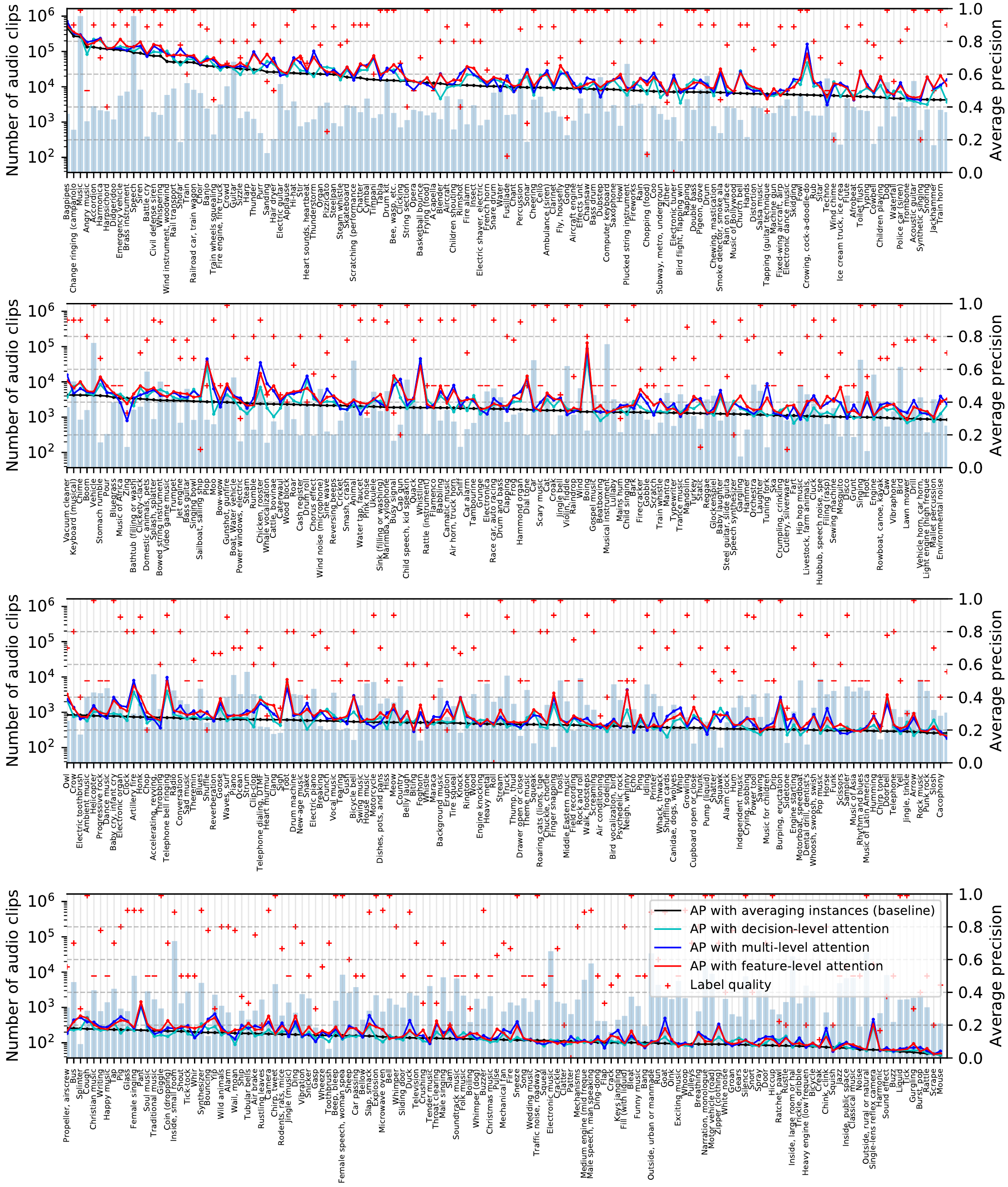}}
  \caption{mAP of all sound classes predicted using different models. }
  \label{fig:full}
\end{figure*}

The results of the feature-level attention neural networks are shown in the bottom block of Table \ref{table:attention}. The ES methods with average and maximum aggregation achieve an mAP of 0.298 and 0.343, respectively. The feature-level attention neural network achieves an mAP of 0.361, an mAUC of 0.969 and a d-prime of 2.641, outperforming the other models. One explanation is that the feature-level attention neural network can attend to or ignore the features in the feature space which further improves the capacity of the decision-level attention neural network. Fig. \ref{fig:aggregation} shows the class-wise performance of the attention neural networks. The feature-level attention neural network outperforms the decision-level attention neural network and the ES method with averaged instances in a majority of sound classes. The results of all 527 sound classes are shown in Fig. \ref{fig:full}.

\begin{table}[t!]
\centering
\caption{Results of decision-level attention model and feature-level attention model}
\label{table:attention}
\begin{tabular}{*{4}{c}}
 \toprule
 & mAP & AUC & d-prime \\
 \midrule
 Average instances \cite{dong2006comparison} & 0.317 & 0.963 & 2.529 \\
 \midrule
 JDC \cite{kong2017joint} & 0.337 & 0.963 & 2.526 \\
 Self attention \cite{ilse2018attention} & 0.324 & 0.962 & 2.506 \\
 \midrule
 Decision-level single-attention \cite{kong2017audio} & 0.337 & 0.968 & 2.612 \\
 Decision-level multi-attention \cite{yu2018multi} & 0.357 & 0.968 & 2.621 \\
 \midrule
 Feature-level avg. pooling & 0.298 & 0.960 & 2.475 \\
 Feature-level max pooling & 0.343 & 0.966 & 2.589 \\
 Feature-level attention & \textbf{0.361} & \textbf{0.969} & \textbf{2.641} \\
 \bottomrule
\end{tabular}
\end{table}

\subsection{Modeling attention function with different functions}
As described in Section \ref{section:exp_different_att_func}, we model the attention function $ q $ of the feature-level attention neural network via a non-negative function $ \phi_{2} $. The choice of the non-negative function may affect the optimisation and result of the attention neural network. Table \ref{table:att_func} shows that the exponential, sigmoid, softmax and NIN functions achieve a similar mAP of approximately 0.360. Modeling $ \phi(\cdot) $ with ReLU is worse than with other non-linear functions.. 

\begin{table}[t!]
\centering
\caption{Results of modeling the non-negative $ \phi_{2} $ with different non-negative functions. }
\label{table:att_func}
\begin{tabular}{*{4}{c}}
 \toprule
 & mAP & AUC & d-prime \\
  \midrule
ReLU att & 0.308 & 0.963 & 2.520 \\
Exp. att & 0.358 & 0.969 & 2.631 \\
Sigmoid att & \textbf{0.361} & \textbf{0.969} & \textbf{2.641} \\
Softmax att & 0.360 & 0.969 & 2.636 \\
NIN & 0.359 & 0.969 & 2.637 \\
 \bottomrule
\end{tabular}
\end{table}

\subsection{Attention neural networks with different embedding depth and width}\label{section:exp_depth}

As shown in (\ref{eq:att_fc_func}), our attention neural networks map the instances $ \mathbf{x} $ to $ \mathbf{x}' $ through several non-linear embedding layers to increase the representation ability of the instances. We model $ f_{\text{FC}} $ using the feature-level attention neural network with fully-connected layers with different depths. Table \ref{table:layers} shows that the mAP increases from 0 layers and reaches a peak of 0.361 at 3 layers. More hidden layers do not increase the mAP. The reason might be that the AudioSet bottleneck features obtained by a VGGish CNN trained on YouTube100M have good separability. Therefore, there is no need to apply very deep neural networks on the AudioSet bottleneck features. On the other hand, the YouTube100M data may have a different distribution from AudioSet. As a result, the embedding mapping $ f_{\text{FC}} $ can be used as domain adaption. 

\begin{table}[t!]
\centering
\caption{Results of modeling the attention neural network with different layer depths. }
\label{table:layers}
\begin{tabular}{*{4}{c}}
  \toprule
  Depth & mAP & AUC & d-prime \\
  \midrule
  \(0\) & 0.328 & 0.963 & 2.522 \\
  \(1\) & 0.356 & 0.967 & 2.605 \\
  \(2\) & 0.358 & 0.968 & 2.620 \\
  \(3\) & \textbf{0.361} & \textbf{0.969} & \textbf{2.641} \\
  \(4\) & 0.356 & 0.969 & 2.637 \\
  \(6\) & 0.348 & 0.968 & 2.619 \\
  \(8\) & 0.339 & 0.967 & 2.595 \\
  \(10\) & 0.331 & 0.966 & 2.579 \\
  \bottomrule
\end{tabular}
\end{table}

Based on the network $ f_{\text{FC}} $ modelled with three layers in the feature-level attention neural network, we investigate the width of $ f_{\text{FC}} $. Table \ref{table:width} shows that feature-level attention model with 2048 hidden units in each hidden layer achieves an mAP of 0.369, an mAUC of 0.969 and a d-prime of 2.641 is achieved, outperforming the models with 256, 512, 1024 and 4096 hidden units in each layer. On the other hand, with 4096 hidden units, the model tends to overfit, and does not outperform the model with 2048 hidden units.

\begin{table}[t!]
\centering
\caption{Results of modeling the attention neural network with different number of hidden units. }
\label{table:width}
\begin{tabular}{*{4}{c}}
  \toprule
  Hidden units & mAP & AUC & d-prime \\
  \midrule
  \(256\) & 0.305 & 0.962 & 2.512 \\
  \(512\) & 0.339 & 0.967 & 2.599 \\
  \(1024\) & 0.361 & 0.969 & \textbf{2.641} \\
  \(2048\) & \textbf{0.369} & \textbf{0.969} & 2.640 \\
  \(4096\) & 0.369 & 0.968 & 2.619 \\
  \bottomrule
\end{tabular}
\end{table}

\section{Conclusion}
We have presented a decision-level and a feature-level attention neural network for AudioSet tagging. We developed the connection between multiple instance learning and attention neural networks. We investigated the class-wise performance of all the 527 sound classes in AudioSet and discovered that the AudioSet tagging performance on AudioSet embedding features is only weakly correlated with the number of training examples and quality of labels, with Pearson correlation coefficients of 0.169 and 0.230, respectively. In addition, we investigated modelling the attention neural networks with different attention functions, depths and widths. Our proposed feature-level attention neural network achieves a state-of-the-art mean average precision (mAP) of 0.369 compared to the best MIL method of 0.317 and the decision-level attention neural network of 0.337. In the future, we will explore weakly labelled sound event detection on AudioSet with attention neural networks.

\section*{Acknowledgment}
The authors would like to thank all anonymous reviewers for their suggestions to improve this paper. 

\ifCLASSOPTIONcaptionsoff
  \newpage
\fi

\bibliographystyle{IEEEtran}	
\bibliography{refs}

\begin{IEEEbiography}[{\includegraphics[width=1in,height=1.25in,clip,keepaspectratio]{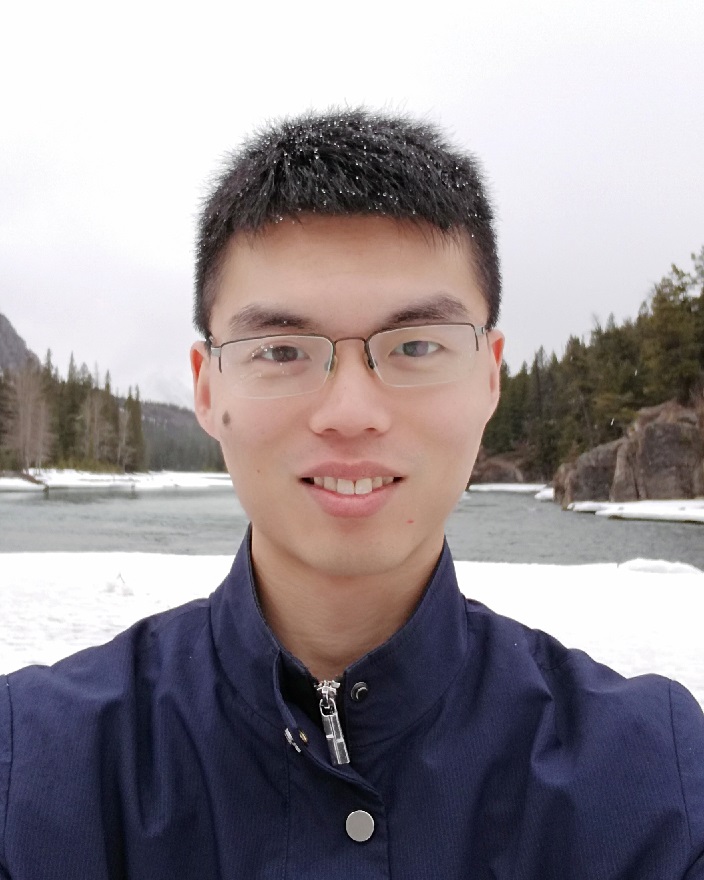}}]{Qiuqiang Kong}
(S'17) received the B.Sc. and M.E. degrees from South China University of Technology, Guangzhou, China, in 2012 and 2015, respectively. He is currently working toward the Ph.D. degree from the University of Surrey, Guildford, U.K on sound event detection. His research topic includes sound understanding, audio signal processing and machine learning. He was nominated as the postgraduate research student of the year in University of Surrey, 2019. 
\end{IEEEbiography}

\vskip 0pt plus -1fil

\begin{IEEEbiography}[{\includegraphics[width=1in,height=1.25in,clip,keepaspectratio]{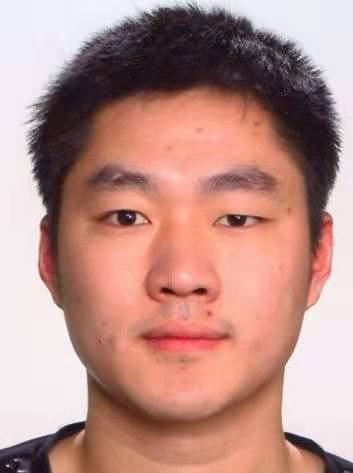}}]{Changsong Yu} received the B.E. degree from Anhalt University of Applied Sciences and M.S. degree University of Stuttgart, Germany, in 2015 and 2018, respectively. He is currently working as simultaneous localization and mapping (SLAM) algorithm engineer in HoloMatic, Beijing, China. His research interest includes deep learning and SLAM.
\end{IEEEbiography}

\vskip 0pt plus -1fil

\begin{IEEEbiography}[{\includegraphics[width=1in,height=1.25in,clip,keepaspectratio]{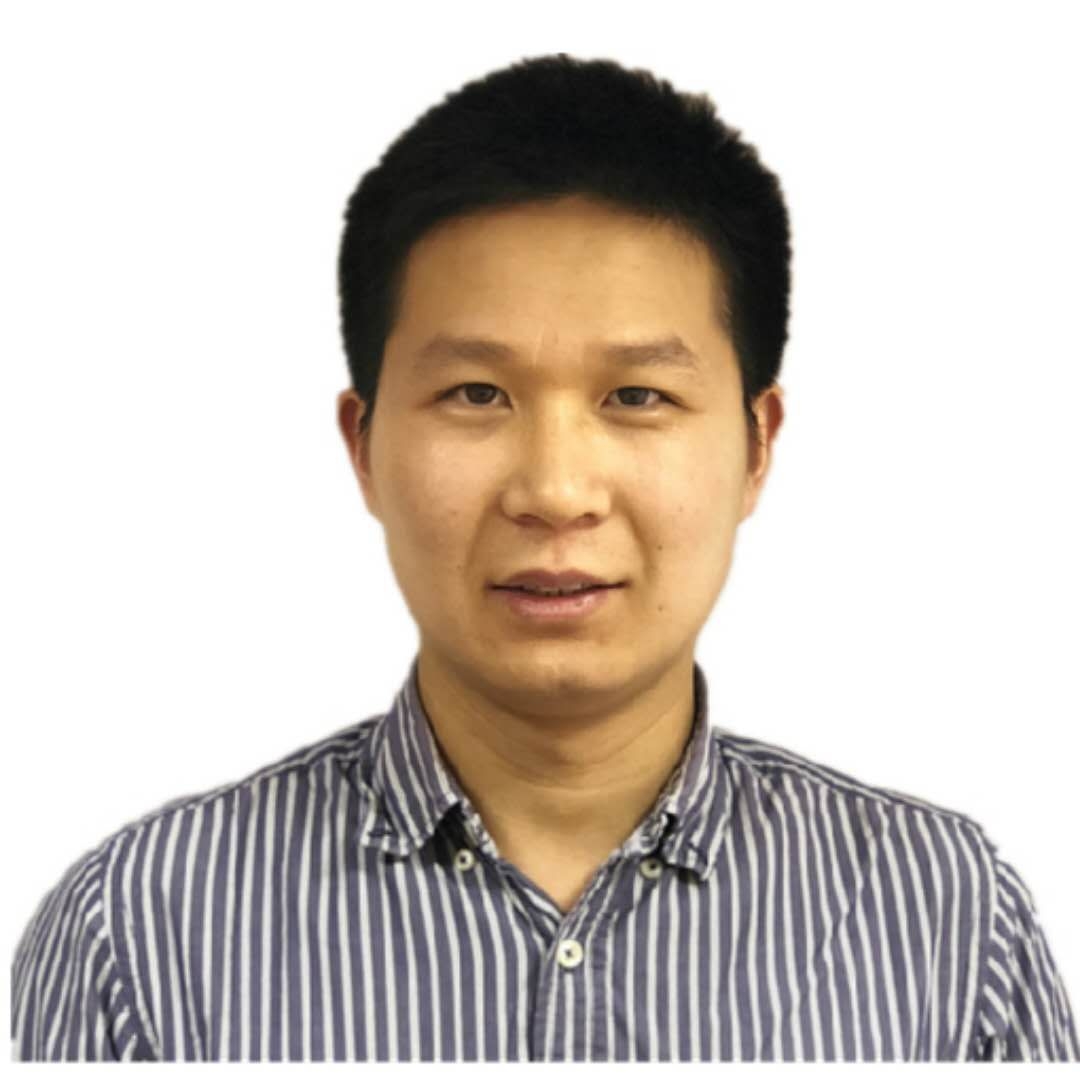}}]{Yong Xu} (M'17) received the Ph.D. degree from the University of Science and Technology of China (USTC), Hefei, China, in 2015, on the topic of DNN-based speech enhancement and recognition. Currently, he is a senior research scientist in Tencent AI lab, Bellevue, USA.  He once worked at the University of Surrey, U.K. as a Research Fellow from 2016 to 2018 working on sound event detection. He visited Prof. Chin-Hui Lee's lab in Georgia Institute of Technology, USA from Sept. 2014 to May 2015. He once also worked in IFLYTEK company from 2015 to 2016 to develop far-field ASR technologies. His research interests include deep learning, speech enhancement and recognition, sound event detection, etc. He received 2018 IEEE SPS best paper award.
\end{IEEEbiography}

\vskip 0pt plus -1fil

\begin{IEEEbiography}[{\includegraphics[width=1in,height=1.25in,clip,keepaspectratio]{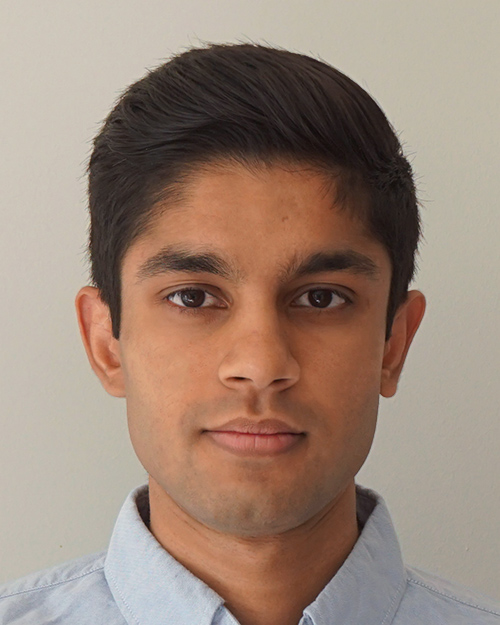}}]{Turab Iqbal} received the B.Eng. degree in Electronic Engineering from the University of Surrey, U.K., in 2017. Currently, he is working towards a Ph.D.  degree from the Centre for Vision, Speech and Signal Processing (CVSSP) in the University of Surrey. His research interests are mainly in machine learning using weakly labeled data for audio classification and localization.
\end{IEEEbiography}

\vskip 0pt plus -1fil

\begin{IEEEbiography}[{\includegraphics[width=1in,height=1.25in,clip,keepaspectratio]{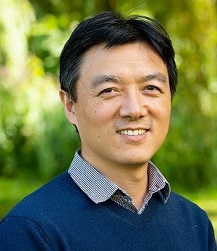}}]{Wenwu Wang} (M'02-SM'11) was born in Anhui, China. He received the B.Sc. degree in 1997, the M.E. degree in 2000, and the Ph.D. degree in 2002, all from Harbin Engineering University, China. He then worked in King's College London, Cardiff University, Tao Group Ltd. (now Antix Labs Ltd.), and Creative Labs, before joining University of Surrey, UK, in May 2007, where he is currently a professor in signal processing and machine learning, and a Co-Director of the Machine Audition Lab within the Centre for Vision Speech and Signal Processing. He has been a Guest Professor at Qingdao University of Science and Technology, China, since 2018. His current research interests include blind signal processing, sparse signal processing, audio-visual signal processing, machine learning and perception, machine audition (listening), and statistical anomaly detection. He has (co)-authored over 200 publications in these areas. He served as an Associate Editor for IEEE TRANSACTIONS ON SIGNAL PROCESSING from 2014 to 2018. He is also Publication Co-Chair for ICASSP 2019, Brighton, UK. 
\end{IEEEbiography}

\vskip 0pt plus -1fil

\begin{IEEEbiography}[{\includegraphics[width=1in,height=1.25in,clip,keepaspectratio]{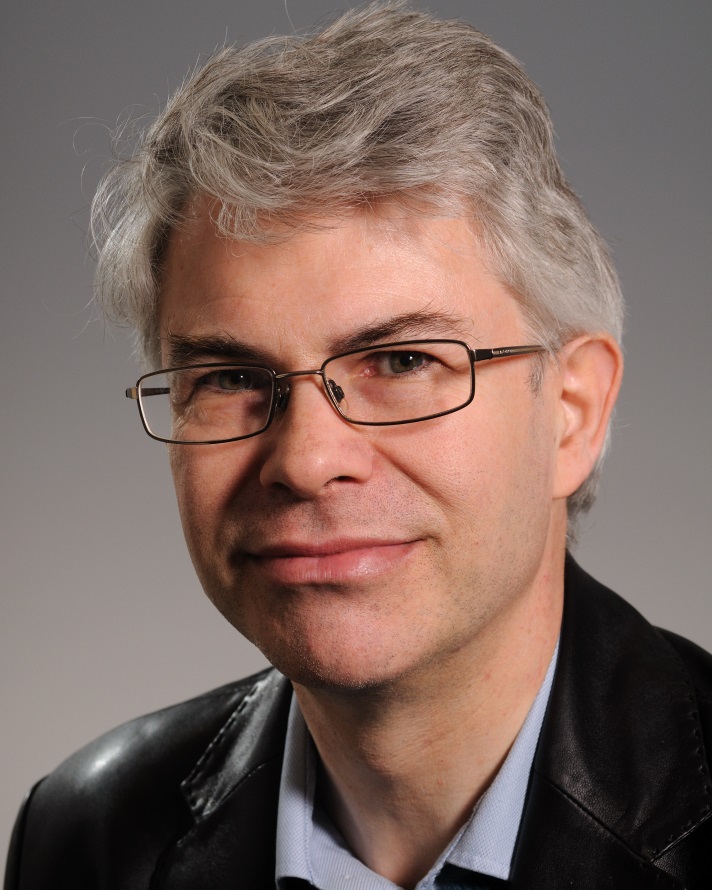}}]{Mark D. Plumbley}
(S'88-M'90-SM'12-F'15) received the B.A.(Hons.) degree in electrical sciences and the Ph.D. degree in neural networks from University of Cambridge, Cambridge, U.K., in 1984 and 1991, respectively. Following his PhD, he became a Lecturer at King's College London, before moving to Queen Mary University of London in 2002. He subsequently became Professor and Director of the Centre for Digital Music, before joining the University of Surrey in 2015 as Professor of Signal Processing. He is known for his work on analysis and processing of audio and music, using a wide range of signal processing techniques, including matrix factorization, sparse representations, and deep learning. He is a co-editor of the recent book on 
Computational Analysis of Sound Scenes and Events,
and Co-Chair of the recent DCASE 2018 Workshop on Detection and Classifications of Acoustic Scenes and Events. He is a Member of the IEEE Signal Processing Society Technical Committee on Signal Processing Theory and Methods, and a Fellow of the IET and IEEE.

\end{IEEEbiography}

\end{document}